\documentclass[prd,aps,tightenlines,superscriptaddress,showpacs,nofootinbib,eqsecnum,amsfonts,amsmath]{revtex4}

\usepackage{epsfig}
\usepackage{graphics}
\usepackage{graphicx}
\usepackage{bm}

\begin{document}

\title{
Complete adiabatic  waveform templates for a test-mass 
in the Schwarzschild spacetime: VIRGO and Advanced LIGO studies
}

\author{P. Ajith}\email{Ajith.Parameswaran@aei.mpg.de}
\affiliation{Max-Planck-Institut f\"ur Gravitationsphysik, 
Albert-Einstein-Institut, Callinstr.38, 30167 Hannover, Germany}
\affiliation{Raman Research Institute, Bangalore 560 080, India}
\author{Bala R. Iyer}\email{bri@rri.res.in} 
\affiliation{Raman Research Institute, Bangalore 560 080, India}
\author{C. A. K. Robinson}\email{Craig.Robinson@astro.cf.ac.uk}
\affiliation{School of Physics and Astronomy,Cardiff University,
5, The Parade, Cardiff, CF24 3YB, U.K.}
\author{B. S. Sathyaprakash}\email{B.Sathyaprakash@astro.cf.ac.uk}
\affiliation{School of Physics and Astronomy,Cardiff University, 
5, The Parade, Cardiff, CF24 3YB, U.K.}

\date{\today}

\begin{abstract}
Post-Newtonian expansions of the binding energy and gravitational wave flux truncated 
at the {\it same relative} post-Newtonian order form the basis of the {\it standard 
adiabatic} approximation to the  phasing of gravitational waves from inspiralling 
compact binaries. Viewed in terms of the dynamics of the binary, the  standard 
approximation is equivalent to neglecting certain conservative post-Newtonian terms 
in the acceleration. In an earlier work, we had proposed a new {\it complete adiabatic} 
approximant constructed from the energy and flux functions. At the leading order it 
employs the 2PN energy function rather than the 0PN one in the standard approximation, 
so that, effectively the approximation corresponds to the dynamics where there are no 
missing post-Newtonian terms in the acceleration. In this paper, we compare the overlaps 
of the standard and complete adiabatic templates with the exact waveform in the adiabatic 
approximation of a test-mass motion in the Schwarzschild spacetime, for the VIRGO and the 
Advanced LIGO noise spectra. It is found that the complete adiabatic approximants 
lead to a remarkable improvement in the {\it effectualness} at lower PN ($<$ 3PN) orders, 
while standard approximants of order $\geq$ 3PN provide a good lower-bound to the complete 
approximants for the construction of effectual templates. {\it Faithfulness} of complete 
approximants is better than that of standard approximants except for a few post-Newtonian 
orders. Standard and complete approximants beyond the adiabatic approximation are also studied using 
the Lagrangian templates of Buonanno, Chen and Vallisneri.
\end{abstract}


\maketitle

\section {Introduction}
Coalescing compact binaries consisting of black holes and/or neutron stars are 
among the most promising sources for ground-based interferometric gravitational
wave (GW) detectors. One of the main challenges of the data analysis for this kind of sources
is to create a template bank with which the detector output may be optimally filtered.
This, in turn, requires an accurate description of the time-evolution of the GW phase. 
Binary coalescences are the end state of a long period of adiabatic dynamics in which 
the orbital frequency of the system changes as a result of gravitational radiation reaction 
but the change in frequency per orbit is negligible compared to the orbital frequency itself.
Then the inspiral orbit can be thought of as an adiabatic perturbation of a number of
circular orbits (with a specific \emph{conserved} energy associated with each of them).
Given the binding energy $E(v)$ and gravitational wave luminosity ${\cal F}(v)$ of 
the binary, the phasing $\varphi(t)$ of the waves can be calculated in the \emph{adiabatic}
approximation using the following ordinary, coupled differential equations:

\begin{equation}
\frac{d\varphi}{dt} = \frac{2v^3}{m},\ \ \ \ 
\frac{dv}{dt} = -\frac{{\cal F}(v)}{mE'(v)},
\label{eq:phasing1}
\end{equation}
where $E'(v)=dE(v)/dv$, and $m=m_1+m_2$ is the total mass of the binary. The binding energy 
$E(v)$ and gravitational wave luminosity ${\cal F}(v)$ are calculated, in general, as 
post-Newtonian (PN) expansions in terms of an invariantly defined velocity parameter $v$
\footnote{Throughout this paper, we use geometrical units in which $G$ = $c$ = 1.}.
The phasing of GWs obtained by numerically solving the above phasing formula
is called \emph{TaylorT1} approximant \cite{dis03}. 

\subsection{Standard and complete approximants}
The standard approach to the GW phasing is based on the PN expansions of the binding 
energy (energy function) and GW luminosity (flux function) truncated at the 
\emph{same relative} PN order \cite{CF}. At the lowest order, it uses only the leading terms in the 
energy (Newtonian) and flux (quadrupolar) functions. For higher order phasing, 
the energy and flux functions are retained to the same relative PN orders. We refer to 
this usual physical treatment of the phasing of GWs computed in the adiabatic 
approximation as the {\it standard adiabatic} approximation. 

With a view to go beyond the adiabatic approximation, we must think of 
this in terms of the dynamics of the binary under conservative relativistic 
forces and gravitational radiation damping. In the conservative dynamics of 
the binary, wherein there is no dissipation, the energy is expressed as 
a post-Newtonian expansion in $v^2$, with the dynamics involving only 
even powers of $v$. When radiation reaction is added to the 
dynamics, then the equation of motion will have terms of both odd and even 
powers of $v$. The radiation reaction is a correction to the dynamics 
that first appears at the 2.5PN (i.e. $v^5$) order 
and not at 1PN or 1.5PN order. It is possible to construct the phasing
of GWs from the direct integration of the equations of motion, without relying
on the adiabatic approximation. At leading order, the \emph{standard non-adiabatic}
approximant uses the 0PN and 2.5PN terms in the acceleration (equivalent to using 
Newtonian conserved energy and quadrupolar flux), neglecting the intervening 1PN 
and 2PN terms. But a complete treatment of the acceleration at leading order 
should include \emph{all} terms up to 2.5PN, without any gaps. We define the 
phasing of the GWs constructed in this approximation as the \emph{complete 
non-adiabatic} approximant.
 
In the \emph{adiabatic} approximation, the energy/flux functions can be thought of
as carrying the information of the conservative/radiative dynamics of the
system. So it is possible to construct adiabatic variants of 
\emph{standard} and {\it complete non-adiabatic} approximants entirely based on energy and
flux functions. In this sense, \emph{standard adiabatic} approximation is equivalent to 
neglecting certain conservative terms in the acceleration and thus resulting in 
gaps in the dynamics. In Ref. \cite{AIRS}, we have proposed a new \emph{complete adiabatic} 
approximant based on energy and flux functions. At the leading order, it uses the 
2PN energy function instead of the 0PN energy function so that, heuristically, 
no intermediate post-Newtonian terms in the acceleration are missed. Following
the notation used in Ref. \cite{AIRS}, we denote the \emph{standard adiabatic} approximant at
$n$PN order as $T(E_{[n]},{\cal F}_{n})$ and the corresponding \emph{complete adiabatic} 
approximant as $T(E_{[n+2.5]},{\cal F}_{n})$, where $[p]$ denotes the integer  part of $p$.

In Ref. \cite{AIRS}, the performance of the standard and complete approximants 
was evaluated in terms of their \emph{effectualness} and \emph{faithfulness} \cite{dis01}.
We first investigated the problem as a general mathematical
question concerning the nature of PN templates assuming a flat power spectrum for the
detector noise. We then repeated the study using the initial LIGO noise spectrum. 
It was found that, at low ($<$3PN) PN orders the complete approximants bring about 
a remarkable improvement over the standard approximants for the construction of 
\emph{effectual} templates. However, the standard approximants are nearly as good as the 
complete approximants at higher orders. In terms of the faithfulness, it was again 
found that the complete approximants generally performed better than the standard 
approximants. 

In the present study, we assess the relative performances of the standard and 
complete adiabatic approximants for the VIRGO and Advanced LIGO noise spectra, as 
characterized by effectualness and faithfulness. We first perform this for the case 
of test-mass templates by comparing them with an exact waveform calculated in the
adiabatic approximation of a test mass falling in to a Schwarzschild black hole. We 
then explore the extension of the results to the comparable mass case, where the 
approximants are compared with a fiducial exact waveform. We restrict our study to 
the inspiral phase of the coalescing binary, neglecting the plunge and quasi-normal 
ring down phases. 

\subsection{Noise spectra of the interferometers}
The one-sided noise power spectral density (PSD) of VIRGO is given in terms of
a dimensionless frequency $x=f/f_0$ by \cite{dis03}
\begin{equation}
S_h(f) = 3.24 \times 10^{-46} \left [ (6.23 x)^{-5} + 2 x^{-1} + 1 + x^ 2 \right]
\end{equation}
where $f_0=500$ Hz; while the same for the Advanced LIGO reads \cite{CT,AISS}
\begin{equation}
S_h(f) = 10^{-49} \left [x^{-4.14} - 5 x^{-2} + 111 \Big(\frac{1 - x^2 + x^4/2}{1 + x^2/2}\Big)\right]
\end{equation}
where $f_0=215$ Hz. 

\section {Test mass waveforms in the adiabatic approximation}
\label{sec:testmass} 
In the case of a test-particle orbiting a Schwarzschild black hole, the energy 
function $E(v)$ is exactly calculable analytically, while the flux function 
${\cal F}(v)$ is exactly calculable numerically \cite{P95}. We use these
functions to construct the exact waveform in the adiabatic approximation.
The analytical exact energy function can be Taylor-expanded to get the 
approximants of the energy function. In addition, ${\cal F}(v)$ has been 
calculated analytically to 5.5PN order~\cite{TTS97} by black hole perturbation 
theory \cite{ST-LivRev}. These approximants of energy and flux function are used to construct the
approximate templates. In this study, we restrict to TaylorT1 approximants, 
since they do not involve any further re-expansion in the phasing 
formula. Hence there is no ambiguity in constructing the phasing of the 
waveforms using approximants with unequal orders of the energy and
flux functions as required for the complete adiabatic approximant
\footnote{See Sec. I.B of Ref. \cite{AIRS} for a detailed discussion.}. The exact waveform 
is also constructed by the TaylorT1 method. The waveforms (both the exact 
and approximate) are all terminated at $v_{\rm lso} =1/\sqrt{6},$ which corresponds 
to $F_{\rm lso}\simeq86$ Hz for the $(1M_{\odot},50M_{\odot})$ binary and 
$F_{\rm lso}\simeq399$ Hz for the $(1M_{\odot},10M_{\odot})$ binary. 
The lower frequency cut-off of the waveforms is chosen to be $F_{\rm low}= 20$ Hz. 

\subsection{Comparison of standard and complete adiabatic approximants}
The effectualness and faithfulness of various PN templates for two archetypical 
binaries with component masses ($1M_{\odot},10M_{\odot}$) and ($1M_{\odot},50M_{\odot}$) 
are tabulated in Tables \ref{table:Effect-TM} and \ref{table:Faith-TM}. 
All the results are in perfect agreement with the results of our earlier
study \cite{AIRS} using the initial LIGO noise spectrum. Complete adiabatic 
approximants bring about a remarkable improvement in the effectualness for all systems at 
low  PN orders ($<$ 3PN). On the other hand, the difference in effectualness between 
the  standard and complete adiabatic approximants at  orders greater than 
3PN is very small. Thus, we conclude that {\it standard adiabatic approximants of 
order $\geq$ 3PN are as good as the complete adiabatic approximants
for the construction of effectual templates}. This study also confirms our earlier 
observation \cite{AIRS} that complete adiabatic approximants are generally less `biased' 
in estimating the parameters of the binary. 

Faithfulness of complete adiabatic approximants is  generally better 
at all PN orders (even at very high orders) studied, which suggests that the 
complete approximants are closer to the exact solution than the corresponding
standard approximants. But there are some cases of anomalous behavior. 
In the next subsection we will try to understand the
reason for these anomalous cases where the complete approximants perform worse 
than the standard.

\begin{table} 
\caption{Effectualness of {\it standard(S) adiabatic} $T(E_{[n]},{\cal F}_{n})$ and 
{\it complete(C) adiabatic} $T(E_{[n+2.5]},{\cal F}_{n})$ approximants in the test mass 
limit. Percentage biases $\sigma_{m}$ and $\sigma_{\eta}$ in determining parameters 
$m$ and $\eta = m_1 m_2/m^2$ are given in brackets.}
\begin{tabular}{cccccccccc}
\hline
\hline
&\vline&\multicolumn{2}{c}{$(1M_{\odot},10M_{\odot})$}
&\vline&\multicolumn{2}{c}{$(1M_{\odot},50M_{\odot})$}\\
\cline{2-7}
PN Order ($n$)&\vline& {\it S} & {\it C} &\vline&  {\it S} & {\it C}\\ 
\hline
Advanced LIGO\\
\hline
0PN   &\vline& 0.4281 (9.2, 2.6) & 0.8960 (32, 42)   &\vline&  0.6461 (27, 22)   & 0.8099 (48, 54)   \\
1PN   &\vline& 0.3498 (28, 8.9)  & 0.7258 (156, 75)  &\vline&  0.6200 (25, 123)  & 0.7093 (27, 13)   \\
1.5PN &\vline& 0.9010 (48, 49)   & 0.9653 (11, 21)   &\vline&  0.6919 (27, 20)   & 0.9532 (2.0, 8.7) \\
2PN   &\vline& 0.9266 (14, 20)   & 0.9814 (2.6, 4.2) &\vline&  0.8835 (31, 39)   & 0.9833 (6.3, 13)  \\
2.5PN &\vline& 0.8917 (89, 66)   & 0.9913 (26, 31.7) &\vline&  0.6720 (26, 6.2)  & 0.9194 (17, 21)   \\
3PN   &\vline& 0.9913 (0.7, 1.6) & 0.9989 (3.9, 7.3) &\vline&  0.9645 (8.4, 16)  & 0.9740 (1.4, 1.4) \\
3.5PN &\vline& 0.9816 (4.5, 7.4) & 0.9994 (0.4, 0.3) &\vline&  0.9875 (14, 23)   & 0.9987 (2.0, 3.9) \\
4PN   &\vline& 0.9895 (4.2, 7.1) & 0.9970 (3.0, 5.3) &\vline&  0.9967 (9.5, 16)  & 0.9973 (4.4, 6.9) \\
4.5PN &\vline& 0.9965 (2.1, 3.6) & 0.9999 (0.8, 1.6) &\vline&  0.9932 (6.1, 11)  & 1.0000 (0.9, 1.9) \\
5PN   &\vline& 0.9954 (2.9, 5.2) & 0.9977 (2.6, 4.0) &\vline&  0.9986 (5.7, 9.9) & 0.9960 (3.6, 6.2) \\
5.5PN &\vline& 0.9963 (2.8, 4.2) & 0.9983 (2.4, 3.9) &\vline&  0.9989 (5.3, 8.7) & 0.9951 (2.7, 4.5) \\
\hline
VIRGO\\
\hline
0PN   &\vline& 0.3894 (42, 41)   & 0.7256 (0.8, 3.8) &\vline& 0.6004 (50, 25)  & 0.8689 (51, 56)   \\
1PN   &\vline& 0.2956 (11, 6.5)  & 0.6876 (187, 80)  &\vline& 0.5498 (51, 30)  & 0.7217 (52, 28)   \\
1.5PN &\vline& 0.8474 (31, 37)   & 0.9487 (12, 22)   &\vline& 0.7308 (56, 53)  & 0.9619 (1.1, 6.9) \\
2PN   &\vline& 0.8933 (9.9, 15)  & 0.9711 (3.0, 4.6) &\vline& 0.9291 (34, 43)  & 0.9854 (5.4, 12)  \\
2.5PN &\vline& 0.8179 (69, 59)   & 0.9864 (26, 32)   &\vline& 0.6579 (49, 41)  & 0.9446 (19, 23)   \\
3PN   &\vline& 0.9845 (0.6, 1.5) & 0.9970 (3.8, 7.3) &\vline& 0.9697 (7.4, 14) & 0.9818 (1.5, 1.5) \\
3.5PN &\vline& 0.9722 (4.3, 7.2) & 0.9991 (0.4, 0.3) &\vline& 0.9885 (14, 22)  & 0.9980 (1.9, 3.8) \\
4PN   &\vline& 0.9829 (4.1, 7.1) & 0.9955 (2.9, 5.2) &\vline& 0.9971 (9.5, 16) & 0.9973 (4.4, 6.9) \\
4.5PN &\vline& 0.9937 (2.0, 3.5) & 0.9999 (0.8, 1.6) &\vline& 0.9926 (6.0, 11) & 1.0000 (0.9, 1.9) \\
5PN   &\vline& 0.9920 (3.0, 5.3) & 0.9967 (2.6, 4.1) &\vline& 0.9987 (5.7, 10) & 0.9960 (3.5, 6.2) \\
5.5PN &\vline& 0.9932 (2.8, 4.2) & 0.9976 (2.3, 3.8) &\vline& 0.9991 (5.4, 9.7)& 0.9948 (2.8, 4.6) \\
\hline
\hline
\label{table:Effect-TM}
\end{tabular}
\end {table}	

\begin{table} 
\caption{Faithfulness of {\it standard(S) adiabatic} $T(E_{[n]},{\cal F}_{n})$ 
and {\it complete(C) adiabatic} $T(E_{[n+2.5]},{\cal F}_{n})$ approximants in the 
test mass limit.}
\begin{tabular}{ccccccccccccccccccc}
\hline
\hline
&\vline&\multicolumn{5}{c}{Advanced LIGO}&\vline&\multicolumn{5}{c}{VIRGO} \\
\cline{2-13}
&\vline&\multicolumn{2}{c}{$(1M_{\odot},10M_{\odot})$} 
&\vline&\multicolumn{2}{c}{$(1M_{\odot},50M_{\odot})$}
&\vline&\multicolumn{2}{c}{$(1M_{\odot},10M_{\odot})$} 
&\vline&\multicolumn{2}{c}{$(1M_{\odot},50M_{\odot})$} \\
\cline{2-13}
PN Order ($n$)&\vline& {\it S} & {\it C} &\vline&  {\it S} & {\it C} 
&\vline& {\it S} & {\it C} &\vline&  {\it S} & {\it C} \\ 
\hline
0PN   &\vline& 0.1456 & 0.4915 &\vline& 0.1608 & 0.2955 &\vline& 0.1384 & 0.3644 &\vline& 0.1265 & 0.3881\\
1PN   &\vline& 0.0853 & 0.1041 &\vline& 0.1159 & 0.1609 &\vline& 0.0682 & 0.0818 &\vline& 0.0887 & 0.1205\\
1.5PN &\vline& 0.2711 & 0.3063 &\vline& 0.2187 & 0.6735 &\vline& 0.2524 & 0.2348 &\vline& 0.1859 & 0.5783\\
2PN   &\vline& 0.6998 & 0.6140 &\vline& 0.2765 & 0.8403 &\vline& 0.7451 & 0.4617 &\vline& 0.2514 & 0.8597\\
2.5PN &\vline& 0.2143 & 0.2710 &\vline& 0.1961 & 0.3094 &\vline& 0.2003 & 0.2496 &\vline& 0.1612 & 0.2420\\
3PN   &\vline& 0.8889 & 0.5791 &\vline& 0.7252 & 0.6971 &\vline& 0.8339 & 0.5745 &\vline& 0.7978 & 0.6210\\
3.5PN &\vline& 0.7476 & 0.9985 &\vline& 0.3852 & 0.9087 &\vline& 0.7684 & 0.9968 &\vline& 0.3821 & 0.9259\\
4PN   &\vline& 0.7314 & 0.8144 &\vline& 0.4404 & 0.5761 &\vline& 0.7501 & 0.7892 &\vline& 0.4024 & 0.5306\\
4.5PN &\vline& 0.9001 & 0.9718 &\vline& 0.5714 & 0.9078 &\vline& 0.8753 & 0.9595 &\vline& 0.5298 & 0.9132\\
5PN   &\vline& 0.8273 & 0.8518 &\vline& 0.5303 & 0.6166 &\vline& 0.8033 & 0.8232 &\vline& 0.4968 & 0.5617\\
5.5PN &\vline& 0.8376 & 0.8640 &\vline& 0.5563 & 0.6460 &\vline& 0.8124 & 0.8340 &\vline& 0.5147 & 0.5862 \\
\hline
\hline
\label{table:Faith-TM}
\end{tabular}
\end {table}	

\subsection{Understanding the results}
Table \ref{table:anomalous} summarizes the PN orders showing the anomalous behavior 
(i.e. the complete approximants being less faithful than the standard approximants) 
for the different noise spectra studied by us. The best-sensitivity bandwidth 
of each detector is shown in brackets \footnote{It should be noted that there is no 
rigorous definition for the `best-sensitivity' bandwidth. We define it as the bandwidth
where the detector's effective noise amplitude $h=\sqrt{f S_h(f)}$ is within a 
factor of two of its lowest value.}. The left-most column in the 
table shows the flattest noise spectrum and the right-most column shows the narrowest
one. 

\begin{figure*}[t]
\centering \includegraphics[width=5in]{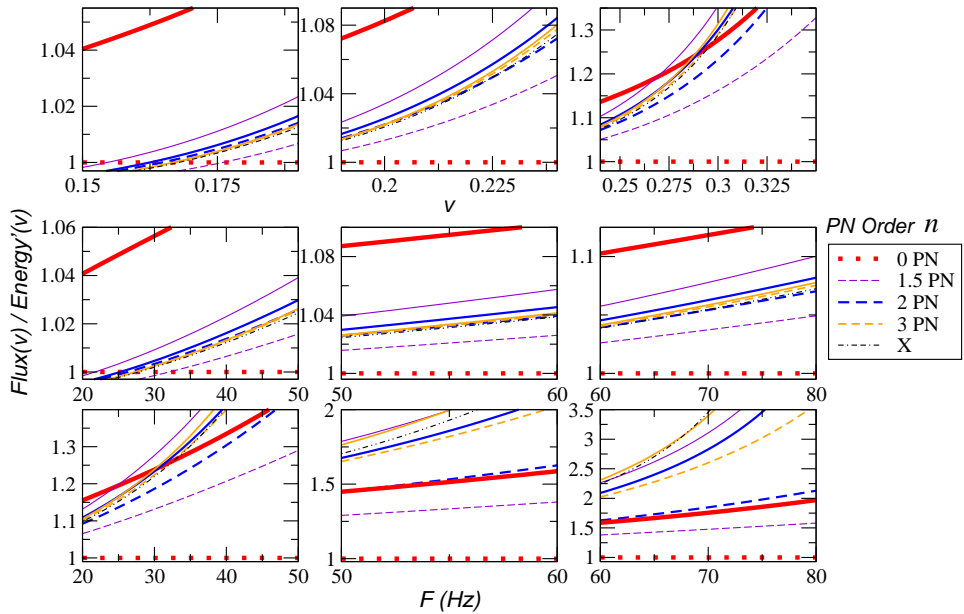}
\caption{Top panel shows the approximants of ${\hat{\cal F}}(v)/\hat{E}'(v)$ plotted as a function 
of $v$. Dashed lines indicate standard approximants ${\hat{\cal F}}_{n}(v)/\hat{E}'_{[n]}(v)$ 
and solid lines indicate the corresponding complete approximants 
${\hat{\cal F}}_{n}(v)/\hat{E}'_{[n+2.5]}(v)$. Middle and bottom panels show the same approximants 
plotted as a function of the GW frequency $F=v^3/\pi m$ in the case of the 
$(1M_\odot,10M_\odot)$ binary and the $(1M_\odot,50M_\odot)$ binary, respectively.}
\label{fig:EbyFofF}
\end{figure*}

In order to understand the anomalous behavior shown at certain PN orders, we
compare the approximants of the ${\cal F}(v)/E'(v)$ function with the corresponding
exact function.  Fig. \ref{fig:EbyFofF} shows the standard and complete approximants of 
${\hat{\cal F}}(v)/\hat{E}'(v)$ along with the exact functions, where a `hat' indicates the
corresponding Newton-normalized quantity. These results show that, while 
the complete approximants are far superior to the standard approximants in modelling 
the late-inspiral, the early inspiral is better modelled by the standard approximants 
at these PN orders. In the case of the $(1M_\odot,10M_\odot)$ binary, the 0PN standard 
approximant is closer to the exact function than the corresponding complete approximant 
in the frequency region 20-50 Hz. But, since none of the detectors is sensitive in this 
frequency band, this effect  shows up in the white-noise case only. Similarly the 1.5PN
and 2PN standard approximants are closer to the exact function in the frequency regions
20-60 Hz and 20-80 Hz, respectively. But the 1.5PN approximant does not show  the 
anomalous behavior in the case of the Advanced LIGO and Initial LIGO because the 20-60 Hz 
region does not fall in the best-sensitivity bandwidth of these detectors. The anomalous 
behavior exhibited by the 3PN approximant can be understood in a similar way. 
It should be noted that the final stages of the inspiral  is  much  better modelled by 
the complete approximants (see the top panel of Fig. \ref{fig:EbyFofF}). But, 
since the binary spends more cycles in the early inspiral phase, the overlaps are 
heavily influenced by the efficiency of the modelling of the early inspiral. 

\begin{table} 
\caption{PN orders showing anomalous behavior in the context of different noise 
spectra. The best-sensitivity bandwidth of each detector is shown in brackets. }
\begin{tabular}{ccccccccccccccccccc}
\hline
\hline
\multicolumn{2}{c}{White-noise}  &\vline&\multicolumn{2}{c}{VIRGO}&\vline&
\multicolumn{2}{c}{Advanced LIGO}&\vline&\multicolumn{2}{c}{Initial LIGO}\\
\multicolumn{2}{c}{}  		 &\vline&\multicolumn{2}{c}{(50-400 Hz)}&\vline&
\multicolumn{2}{c}{(60-300 Hz)}  &\vline&\multicolumn{2}{c}{(80-200 Hz)}\\
\hline
 $(1,10)M_\odot$ & $(1,50)M_\odot$ &\vline& $(1,10)M_\odot$ & $(1,50)M_\odot$&\vline&
 $(1,10)M_\odot$ & $(1,50)M_\odot$ &\vline& $(1,10)M_\odot$ & $(1,50)M_\odot$\\
\hline
0PN	&	&\vline&	&	&\vline&	&	&\vline&     &	\\
1.5PN	&	&\vline& 1.5PN	&	&\vline&	&	&\vline&     &	\\
2PN	&	&\vline& 2PN	&	&\vline& 2PN	&	&\vline&     &	\\
3PN	& 3PN   &\vline& 3PN	& 3PN   &\vline& 3PN	& 3PN   &\vline& 3PN &  \\
\hline
\hline
\label{table:anomalous}
\end{tabular}
\end {table}	

\section {Comparable mass waveforms in the adiabatic approximation}
\label{sec:compmass} 
In the case of comparable mass binaries there is no {\it exact} waveform
available and the best we can do is to compare the performance of the standard 
adiabatic and complete adiabatic templates by studying their overlaps with 
some plausible fiducial exact waveform. The required energy and flux functions 
have been calculated by supplementing the exact functions in the test mass 
limit by all the {\it known}  $\eta$-dependent corrections (from post-Newtonian 
theory~\cite{Luc-LivRev}) in the comparable mass case\footnote{Sec. IV of Ref. \cite{AIRS} discusses 
the precise method used in this construction.}. In the case of comparable mass binaries, 
the energy function is currently known up to 3PN order 
\cite{DJS,BF,DJS02,BDE03,IF,itoh2} and the flux function up to 3.5PN order
\cite{BDIWW,BDI,WW,BIWW,B96,BIJ02,BFIJ02,ABIQ04,BDEI,BI04,BDI04,BDEI05}. 
The waveforms (`exact' and approximate) are constructed by  the TaylorT1 
method and are terminated at $v_{\rm lso}= 1/\sqrt{6}$, which corresponds to 
$F_{\rm lso}\simeq 1570$ Hz for a $(1.4M_{\odot},\, 1.4M_{\odot})$ 
binary and $F_{\rm lso}\simeq 220$ Hz for a $(10M_{\odot},\, 10M_{\odot})$ binary. 
The lower frequency cut-off of the waveforms is chosen to be $F_{\rm low}= 20$ Hz. 

\subsection{Comparable mass results in the adiabatic approximation}
\begin{figure*}[t]
\centering \includegraphics[width=5in]{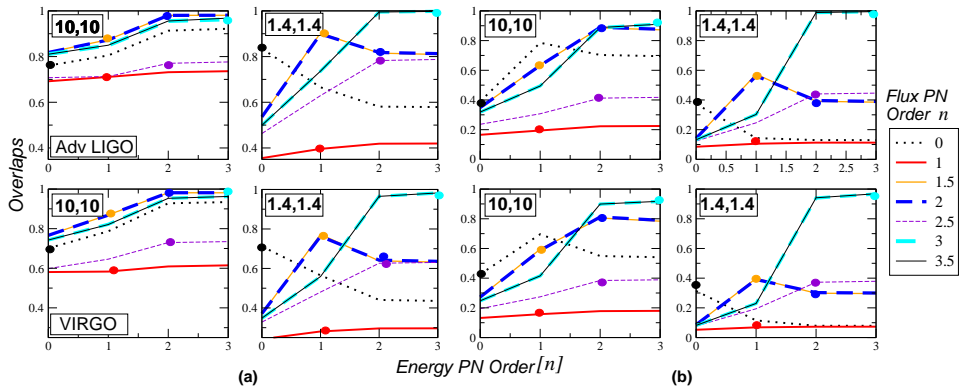}
\caption{Effectualness (four panels in the left) and faithfulness (four panels in 
the right) of various TaylorT1 templates in detecting two prototype binaries with
component masses $(10M_{\odot},\, 10M_{\odot})$ and $(1.4M_{\odot},\, 1.4M_{\odot})$. 
Top panels show overlaps calculated using the Advanced LIGO noise spectrum and the 
bottom panels show overlaps calculated using the VIRGO noise spectrum. Different lines 
in the panels correspond to different orders of the flux function. Each line shows how 
the overlaps are evolving as a function of the accuracy of the energy function. Standard 
adiabatic approximants $T(E_{[n]},{\cal F}_n)$ are marked with thick dots.}
\label{fig:CM-overlaps}
\end{figure*}
Effectualness and faithfulness of various TaylorT1 templates in detecting two 
prototype binaries with component masses $(10M_{\odot},\, 10M_{\odot})$ and 
$(1.4M_{\odot},\, 1.4M_{\odot})$ are shown in Fig. \ref{fig:CM-overlaps} and 
are tabulated in Tables \ref{table:Effect-CM} and \ref{table:Faith-CM}.
It should be noted that the complete adiabatic approximants are presently calculated 
only up to 1PN order. Thus, it is not possible to make strong statements of general 
trends. Heuristically one can conclude from Fig. \ref{fig:CM-overlaps} 
that standard adiabatic approximants of order $\geq$ 1.5PN provide a good 
lower-bound to the complete adiabatic approximants for the construction of both 
effectual and faithful templates. We also note from Table \ref{table:Effect-CM} 
that standard adiabatic approximants of order 2PN/3PN produce the target value 
$0.965$ in effectualness (corresponding to 10\% event-loss) in the case of the 
BH-BH/NS-NS binaries.

\begin{table} 
\caption{Effectualness of {\it standard(S) adiabatic} $T(E_{[n]},{\cal F}_{n})$ and 
{\it complete(C) adiabatic} $T(E_{[n+2.5]},{\cal F}_{n})$ approximants in the comparable-mass 
case. Percentage biases $\sigma_{m}$ and $\sigma_{\eta}$ in determining parameters 
$m$ and $\eta=m_1 m_2/m^2$ are given in brackets.}
\begin{tabular}{ccccccccccccccc}
\hline
\hline
&\vline&\multicolumn{3}{c}{$(10M_{\odot},10M_{\odot})$}
&\vline&\multicolumn{3}{c}{$(1.4M_{\odot},1.4M_{\odot})$}\\
\cline{2-9}
PN Order ($n$)&\vline& {\it S} && {\it C} &\vline&  {\it S} && {\it C}\\ 
\hline
Advanced LIGO \\
\hline
0PN   	&\vline& 0.7606 (8.5, 0.1) && 0.9132 (5.0, 0.3) &\vline& 0.8347 (1.4, 0.1) && 0.5809 (4.3, 0.1)  \\
1PN   	&\vline& 0.7110 (57, 0.6)  && 0.7360 (39, 0.6)  &\vline& 0.3959 (6.1, 0.0) && 0.4194 (5.0, 0.1)  \\
1.5PN 	&\vline& 0.8741 (2.4, 0.1) &&                   &\vline& 0.9034 (0.0, 0.2) &&                   \\
2PN   	&\vline& 0.9803 (0.8, 0.2) &&                   &\vline& 0.8179 (0.4, 0.0) &&                   \\
2.5PN 	&\vline& 0.7705 (7.3, 0.1) &&                   &\vline& 0.7826 (0.4, 0.0) &&                   \\
3PN   	&\vline& 0.9626 (0.5, 0.0) &&                   &\vline& 0.9981 (0.4, 0.3) &&                   \\
3.5PN 	&\vline& 0.9683 (1.3, 1.5) &&                   &\vline& 0.9977 (0.4, 0.3) &&                   \\
\hline
VIRGO \\
\hline
0PN   	&\vline& 0.7009 (5.3, 0.0) && 0.9280 (4.7, 0.9) &\vline& 0.7119 (1.4, 0.1) && 0.4405 (5.0, 0.0) \\
1PN   	&\vline& 0.5834 (56, 0.3)  && 0.6148 (30, 0.2)  &\vline& 0.2808 (3.9, 0.1) && 0.2968 (2.9, 0.0) \\
1.5PN 	&\vline& 0.8698 (1.3, 0.0) &&                   &\vline& 0.7724 (0.4, 0.7) &&                  \\
2PN   	&\vline& 0.9815 (0.8, 0.2) &&                   &\vline& 0.6420 (0.0, 0.0) &&                  \\
2.5PN 	&\vline& 0.7299 (4.8, 0.0) &&                   &\vline& 0.6266 (0.0, 0.1) &&                  \\
3PN   	&\vline& 0.9624 (0.5, 0.1) &&                   &\vline& 0.9822 (0.0, 0.3) &&                  \\
3.5PN 	&\vline& 0.9627 (0.5, 0.1) &&                   &\vline& 0.9823 (0.0, 0.3) &&                  \\
\hline
\hline
\label{table:Effect-CM}
\end{tabular}
\end {table}

\begin{table} 
\caption{Faithfulness of the {\it standard(S) adiabatic} $T(E_{[n]},{\cal F}_{n})$ 
and {\it complete(C) adiabatic} $T(E_{[n+2.5]},{\cal F}_{n})$ approximants in the 
comparable-mass case.}
\begin{tabular}{ccccccccccccccccccc}
\hline
\hline
&\vline&\multicolumn{5}{c}{Advanced LIGO}&\vline&\multicolumn{5}{c}{VIRGO} \\
\cline{2-13}
&\vline&\multicolumn{2}{c}{$(10M_{\odot},10M_{\odot})$} 
&\vline&\multicolumn{2}{c}{$(1.4M_{\odot},1.4M_{\odot})$}
&\vline&\multicolumn{2}{c}{$(10M_{\odot},10M_{\odot})$} 
&\vline&\multicolumn{2}{c}{$(1.4M_{\odot},1.4M_{\odot})$} \\
\cline{2-13}
PN Order ($n$)&\vline& {\it S} & {\it C} &\vline&  {\it S} & {\it C} 
&\vline& {\it S} & {\it C} &\vline&  {\it S} & {\it C} \\ 
\hline
0PN   &\vline& 0.3902 & 0.7030 &\vline& 0.3731 & 0.1300 &\vline& 0.4262 & 0.5490 &\vline& 0.3138 & 0.0794\\
1PN   &\vline& 0.1944 & 0.2248 &\vline& 0.1054 & 0.1128 &\vline& 0.1574 & 0.1798 &\vline& 0.0686 & 0.0732\\
1.5PN &\vline& 0.6362 &        &\vline& 0.5735 &        &\vline& 0.5950 &        &\vline& 0.3986 &       \\
2PN   &\vline& 0.8895 &        &\vline& 0.3964 &        &\vline& 0.8120 &        &\vline& 0.3027 &       \\
2.5PN &\vline& 0.4125 &        &\vline& 0.4407 &        &\vline& 0.3842 &        &\vline& 0.3726 &       \\
3PN   &\vline& 0.9117 &        &\vline& 0.9947 &        &\vline& 0.9169 &        &\vline& 0.9668 &       \\
3.5PN &\vline& 0.9106 &        &\vline& 0.9952 &        &\vline& 0.9177 &        &\vline& 0.9686 &       \\
\hline
\hline
\label{table:Faith-CM}
\end{tabular}
\end {table}
	
\section{Non-adiabatic approximants}
\begin{table} 
\caption{Effectualness and faithfulness of the {\it standard(S)} and {\it complete(C) 
non-adiabatic approximants} in the test mass case.}
\begin{tabular}{ccccccccccccccccc}
\hline
\hline
&\vline&\multicolumn{5}{c}{Effectualness} &\vline&\multicolumn{5}{c}{Faithfulness} \\
\cline{2-13}
&\vline&\multicolumn{2}{c}{$(1M_{\odot},10M_{\odot})$} 
&\vline&\multicolumn{2}{c}{$(1M_{\odot},50M_{\odot})$}
&\vline&\multicolumn{2}{c}{$(1M_{\odot},10M_{\odot})$} 
&\vline&\multicolumn{2}{c}{$(1M_{\odot},50M_{\odot})$} \\
\cline{2-13}
PN Order ($n$)&\vline& {\it S} & {\it C} &\vline&  {\it S} & {\it C} 
&\vline& {\it S} & {\it C} &\vline&  {\it S} & {\it C} \\ 
\hline
Advanced LIGO\\
\hline
0PN & \vline & 0.4259 & 0.8682 & \vline & 0.6384 & 0.9360 & \vline & 0.2636 & 0.0754  & \vline & 0.4857 & 0.0999 \\
1PN & \vline & 0.5256 & 0.8280 & \vline & 0.6080 & 0.9211 & \vline & 0.3419 & 0.1368  & \vline & 0.4493 & 0.2038 \\
\hline
VIRGO\\
\hline
0PN & \vline & 0.3720 & 0.7631 & \vline & 0.5985 & 0.9618 & \vline & 0.1991 & 0.0570  & \vline & 0.4947 & 0.1057 \\
1PN & \vline & 0.3599 & 0.7386 & \vline & 0.5777 & 0.9525 & \vline & 0.2499 & 0.0911  & \vline & 0.5259 & 0.1954 \\
\hline
\hline
\label{table:NonAdb-TM}
\end{tabular}
\end {table}	

\begin{table} 
\caption{Effectualness and faithfulness of the {\it standard(S)} and {\it complete(C) 
non-adiabatic approximants} in the comparable mass case.}
\begin{tabular}{ccccccccccccccccc}
\hline
\hline
&\vline&\multicolumn{5}{c}{Effectualness} &\vline&\multicolumn{5}{c}{Faithfulness} \\
\cline{2-13}
&\vline&\multicolumn{2}{c}{$(10M_{\odot},10M_{\odot})$} 
&\vline&\multicolumn{2}{c}{$(1.4M_{\odot},1.4M_{\odot})$} 
&\vline&\multicolumn{2}{c}{$(10M_{\odot},10M_{\odot})$}
&\vline&\multicolumn{2}{c}{$(1.4M_{\odot},1.4M_{\odot})$} \\
\cline{2-13}
PN Order ($n$)&\vline& {\it S} & {\it C} &\vline&  {\it S} & {\it C} 
&\vline& {\it S} & {\it C} &\vline&  {\it S} & {\it C} \\ 
\hline
Advanced LIGO\\
\hline
0PN    & \vline & 0.9147 & 0.4338 & \vline & 0.7417 & 0.8322 & \vline & 0.0637 & 0.0546 & \vline & 0.4662 & 0.2192\\
1PN    & \vline & 0.3937 & 0.5132 & \vline & 0.7443 & 0.8158 & \vline & 0.0794 & 0.0703 & \vline & 0.6594 & 0.4788\\
\hline
VIRGO\\
\hline
0PN    & \vline & 0.8142 & 0.3113 & \vline & 0.6895 & 0.7341 & \vline & 0.0414 & 0.0334  & \vline & 0.3439 & 0.1661\\
1PN    & \vline & 0.2880 & 0.3944 & \vline & 0.6807 & 0.7420 & \vline & 0.0803 & 0.0463  & \vline & 0.5709 & 0.3704 \\
\hline
\hline
\label{table:NonAdb-CM}
\end{tabular}
\end {table}	
In \cite{AIRS}, we also looked at the relative performance of {\it standard} and {\it complete
non-adiabatic approximants}, using the Lagrangian templates described by 
Buonanno, Chen and Vallisneri \cite{BCV02}, comparing them with the exact waveform 
\footnote{In the comparable mass case, the approximants were compared with a fiducial 
exact waveform which is discussed in Sec. \ref{sec:compmass}.} constructed
in the adiabatic approximation. The overlaps were computed for the white noise spectrum 
and the initial LIGO noise spectrum. We had found that, as for 
the adiabatic approximants, the complete approximants were far
better than the standard approximants for the construction of effectual templates. However,
in marked contrast to the adiabatic case, we found that the use of complete templates led
to a decrease in faithfulness compared to the standard templates.

In this study, we look at the relative performance of the standard and complete approximants
for the case of the Advanced LIGO and VIRGO noise spectra. The results are tabulated in Tables
\ref{table:NonAdb-TM} and \ref{table:NonAdb-CM}. It should be noted that, at present,
results are available at too few PN orders to make statements about general trends in effectualness
and faithfulness. However, we find that, as in \cite{AIRS}, the complete approximants are 
generally better than the standard approximants for the construction of effectual 
templates, but the faithfulness of the complete approximants is less than that of 
the standard.

\section{Summary}
The \emph{standard adiabatic} approximation to the phasing of GWs from inspiralling compact
binaries is based on PN expansions of the binding energy and GW flux truncated at 
the \emph{same relative} PN order. Viewed in terms of the dynamics of the binary,
this standard treatment is equivalent to neglecting certain conservative terms in 
the acceleration. In an earlier work \cite{AIRS}, we have proposed a new \emph{complete
adiabatic} approximant which, in spirit, corresponds to a complete treatment of the 
acceleration accurate up to the respective PN order. In this study we have investigated 
the performance of the \emph{standard} and \emph{complete
adiabatic} approximants in the cases of the VIRGO and Advanced LIGO noise spectra.
This has been done by measuring their \emph{effectualness} (i.e. larger maximum overlap
with the exact waveform), and \emph{faithfulness} (i.e. smaller biases in parameter
estimation). In the test-mass case, the approximants were compared with the exact 
waveform; while in the comparable mass case, the approximants were compared with a fiducial `exact' 
waveform. We have considered only the inspiral phase of the binary, neglecting the 
plunge and quasi-normal ring down phases. The results of this study are in full agreement
with our earlier studies \cite{AIRS} using Initial LIGO noise spectrum and white noise spectrum. 
We summarize the results of our study as follows:

\begin{itemize}

\item In the test-mass case, effectualness of the templates improves significantly 
in the complete adiabatic approximation at lower ($<$ 3PN) PN orders. But standard 
adiabatic approximants of order $\geq$ 3PN are nearly as good as the complete
approximants. 

\item Faithfulness of complete adiabatic approximants is generally better 
at all PN orders studied. But there are some cases of anomalous behavior. 
We have shown that, at these PN orders (0PN, 1.5PN, 2PN and 3PN) the early inspiral
is better modelled by the standard approximants than the corresponding complete
approximants, which explains the better faithfulness exhibited by the standard 
approximants at these orders. But complete adiabatic approximants are far superior
to the standard adiabatic approximants in modelling the final inspiral.  

\item Complete adiabatic approximants are generally less `biased' in estimating the parameters
of the binary. 

\item In the case of comparable-mass binaries, standard adiabatic approximants of 
order $\geq$ 1.5PN provides a good lower-bound to the complete adiabatic approximants
for the construction of both effectual and faithful templates. 

\item In the comparable-mass case, standard adiabatic approximants of order 2PN/3PN 
produce the target value $0.965$ in effectualness (corresponding to 10\% event-loss) 
in the case of the BH-BH/NS-NS binaries.

\end{itemize}

\end {document}